\documentclass[11pt,a4paper,american]{extarticle}
\usepackage[T1]{fontenc}
\usepackage[latin1]{inputenc}
\usepackage{amsmath}
\usepackage{amssymb}
\usepackage{setspace}
\usepackage{esint}
\usepackage[numbers]{natbib}
\makeatletter

\newenvironment{lyxlist}[1]
{\begin{list}{}
{\settowidth{\labelwidth}{#1}
 \setlength{\leftmargin}{\labelwidth}
 \addtolength{\leftmargin}{\labelsep}
 }}
{\end{list}}

\usepackage{babel}
\usepackage{bbm}
\usepackage{hyperref}
\usepackage{authblk}
\usepackage{tikz}
\usetikzlibrary{matrix,shapes.geometric,arrows}
\usepackage{cancel}
\date{}
\allowdisplaybreaks
\numberwithin{equation}{section}

\author{Renann Lipinski Jusinskas\thanks{renannlj@fzu.cz}}
\affil{Institute of Physics of the Czech Academy of Sciences \\ CEICO - Central European Institute for Cosmology and Fundamental Physics\authorcr  Na Slovance 2, 182 21, Prague - Czech Republic}

\makeatother

\usepackage{babel}
\begin{document}

\title{Quantization of the particle with a linear massless solution}
\maketitle
\begin{abstract}
In this work, a solution linear in the momentum for the massless constraint
$P^{m}P_{m}=0$ is investigated. It is presented in terms of a $SO(2n,\mathbb{C})$
to $U(n)$ decomposition and interpreted in terms of projective pure
spinors, which are known to parametrize the $\frac{SO(2n)}{U(n)}$
coset. The worldline action is quantized using the BRST formalism
and, using the results of Berkovits and Cherkis, the ghost number
zero wave function is shown to generate massless solutions for field
equations of arbitrary spin. The model can be covariantly expressed
by the action recently proposed in $D=10$ by Berkovits, in terms
of a twistor-like constraint. However, a thorough account of its gauge
symmetries does not lead to a spacetime supersymmetric theory. In
order to derive from first principles the superparticle in the pure
spinor formalism, a new model is proposed with partial worldline supersymmetry.
The gauge algebra is then analyzed within the Batalin-Vilkovisky formalism
and the gauge fixed action is finally shown to describe the pure spinor
superparticle times a $U(1)$ decoupled sector.\tableofcontents{}

\pagebreak{}
\end{abstract}

\section{Overview}

The worldline action for a massless particle in Minkowski space (metric
$\eta^{mn}$) is given by
\begin{equation}
S_{p}=\int d\tau\{P_{m}\dot{X}^{m}-\tfrac{1}{2}e\eta^{mn}P_{m}P_{n}\},\label{eq:action1st}
\end{equation}
where $X^{m}(\tau)$ denotes the coordinates of the particle at a
given instant $\tau$, $\dot{X}^{m}\equiv\frac{dX^{m}}{d\tau}$, $P_{m}$
stands for the conjugate momentum of $X^{m}$ and the \emph{einbein}
$e$ is the Lagrange multiplier imposing the mass-shell. By construction,
$S_{p}$ has a gauge symmetry characterized by the invariance under
reparametrization of $\tau$.

Quantization of the action \eqref{eq:action1st} leads to a scalar
field $\varphi$ satisfying the wave equation. Within the BRST formalism,
$\varphi$ appears as the ghost number zero cohomology of the BRST
charge
\begin{equation}
Q=cP^{m}P_{m},
\end{equation}
where $(b,c)$ is the ghost conjugate pair associated to reparametrization
symmetry and the gauge fixed action takes the form
\begin{equation}
S=\int d\tau\{P_{m}\dot{X}^{m}-\tfrac{1}{2}P_{m}P^{m}+b\dot{c}\}.
\end{equation}
Note that the choice $e=0$ is not physically sensible as it corresponds
to a degenerate worldline metric.

Understanding the massless particle is particularly interesting because
it can be viewed in some sense as the zero length limit of the string.
Therefore, alternative formulations of the particle action might provide
some insight to develop different string descriptions. The quantization
of the particle has been extensively explored in the past (see \cite{Sorokin:1999jx}
and references therein for a review). Among the different techniques,
the twistor parametrization introduced by Penrose in $D=4$ \cite{Penrose:1967wn}
is one of the most fruitful. In \cite{Bars:2005ze}, for example,
the twistor description was extended from massless to massive particles,
including $AdS$ backgrounds. 

In this work, the quantization of the massless particle will be investigated
with a particular solution for the mass-shell condition. In order
to do so, the Lorentz group $SO(2n-1,1)$ will be described in terms
of a complexified $SO(2n)$ group and the worldline fields will be
presented in terms of $U(n)$ representations. For $X^{m}$ and $P_{m}$
the decomposition is\begin{subequations}
\begin{eqnarray}
X^{m} & = & \{x^{a},\bar{x}_{a}\},\\
P_{m} & = & \{p_{a},\bar{p}^{a}\},
\end{eqnarray}
\end{subequations}with $m=1,\ldots,2n$ and $a=1,\ldots,n$, such
that
\begin{equation}
\begin{array}{rclcrcl}
x^{a} & \equiv & \frac{1}{\sqrt{2}}(X^{2a-1}+iX^{2a}), &  & p_{a} & \equiv & \frac{1}{\sqrt{2}}(P_{2a-1}-iP_{2a}),\\
\bar{x}_{a} & \equiv & \frac{1}{\sqrt{2}}(X^{2a-1}-iX^{2a}), &  & \bar{p}^{a} & \equiv & \frac{1}{\sqrt{2}}(P_{2a-1}+iP_{2a}),
\end{array}\label{eq:vectordecomposition}
\end{equation}
and the massless condition $P^{m}P_{m}=0$ is rewritten as
\begin{equation}
\bar{p}^{a}p_{a}=0.\label{eq:MasslessCondition}
\end{equation}

Any solution of \eqref{eq:MasslessCondition} in $D=2n$ dimensions
can be put in the form
\begin{equation}
\bar{p}^{a}+\gamma^{ab}p_{b}=0,\label{eq:MasslessSolution}
\end{equation}
for some antisymmetric $U(n)$ tensor $\gamma^{ab}$. As will be shown
here, \eqref{eq:MasslessSolution} is the irreducible part of the
twistor-like constraint
\begin{equation}
(\gamma^{m}\lambda)_{\alpha}P_{m}=0,\label{eq:BerkovitsTwistorLike}
\end{equation}
proposed in \cite{Berkovits:2011gh}, where $\gamma^{m}$ denotes
a chiral block of the Dirac matrices and $\lambda^{\alpha}$ is a
projective pure spinor, with $\alpha=1,\ldots,2^{n-1}$. Therefore,
$\gamma^{ab}$ can be interpreted in terms of projective pure spinors,
known to parametrize the $\frac{SO(2n)}{U(n)}$ coset \cite{Nekrasov:2005wg}. 

In \cite{Hughston}, Hughston showed that solutions to the wave equation
in even dimensions can be constructed using projective pure spinors.
Later on, this idea was extended with the proper definition of the
projective pure spinors integration measure and the explicit construction
of solutions of massless field equations of arbitrary spin \cite{Berkovits:2004bw}.

As it turns out, the BRST quantization of the massless particle action
subjected to the constraint \eqref{eq:MasslessSolution} gives rise
to a similar structure and its ghost number zero cohomology has a
simple form which can be expressed by the wave function
\begin{equation}
F=F(\mathbb{X}^{a},\gamma^{ab}).\label{eq:masslessarbitraryspin}
\end{equation}
Here, $\mathbb{X}^{a}=x^{a}+\gamma^{ab}\bar{x}_{b}$ corresponds to
the independent components of the spinor $\mathbb{X}_{\alpha}$ defined
by
\begin{equation}
\mathbb{X}_{\alpha}\equiv X^{m}(\gamma_{m}\lambda)_{\alpha}.
\end{equation}
In \cite{Berkovits:2004bw}, it was shown that $F$ is the generating
function of massless solutions of field equations of spin $\tfrac{N}{2}$,
given by 
\begin{equation}
\phi^{(\alpha_{1}\cdots\alpha_{N})}=\int[d\lambda]\lambda^{\alpha_{1}}\ldots\lambda^{\alpha_{N}}F(\mathbb{X}_{\alpha},\lambda^{\alpha}),\label{eq:masslesssolutions}
\end{equation}
where $[d\lambda]$ is some integration measure over the projective
pure spinors.

More recently \cite{Berkovits:2015yra}, the constraint \eqref{eq:BerkovitsTwistorLike}
was implemented in an effort to provide a bona fide gauge structure
that would finally lead to the pure spinor superparticle \cite{Berkovits:2001rb}.
However, the wave function \eqref{eq:masslessarbitraryspin} encodes
a field content far richer than the physical spectrum of the superparticle.
In addition, a more extensive analysis of the gauge symmetries of
the worldline action proposed in \cite{Berkovits:2015yra} does not
lead to a spacetime supersymmetric theory.

On the other hand, the analysis of the physical degrees of freedom
of the ten-dimensional model promptly hints at its generalization.
The new ingredients are the anticommuting spinor $\bar{\theta}^{\alpha}$,
a ``superpartner'' for the pure spinor $\lambda^{\alpha}$, satisfying
\begin{equation}
(\lambda\gamma^{m}\bar{\theta})=0,
\end{equation}
with conjugate $\bar{p}_{\alpha}$, and a sort of local worldline
supersymmetry for the spacetime spinors, generated by $\lambda^{\alpha}\bar{p}_{\alpha}$.
The gauge fixed action of this model and its associated BRST charge
can be cast as\begin{subequations}
\begin{eqnarray}
S_{\tiny{PS}} & = & \int d\tau\{P_{m}\dot{X}^{m}+w_{\alpha}\dot{\lambda}^{\alpha}-\tfrac{1}{2}P_{m}P^{m}+\bar{p}_{\alpha}\dot{\bar{\theta}}^{\alpha}+\hat{p}_{\alpha}\dot{\hat{\theta}}^{\alpha}+\bar{\Omega}\dot{\Omega}+\beta\dot{\gamma}\},\label{eq:FINALPS}\\
Q & = & \gamma\lambda^{\alpha}\bar{p}_{\alpha}-(\lambda\gamma^{m}\hat{\theta})P_{m}+\Omega(\beta\gamma-\hat{\theta}^{\alpha}\hat{p}_{\alpha}-\lambda^{\alpha}w_{\alpha}).
\end{eqnarray}
\end{subequations}Here, the conjugate pairs $\{\hat{p}_{\alpha},\hat{\theta}^{\alpha}\}$,
$\{\bar{\Omega},\Omega\}$ and $\{\beta,\gamma\}$ respectively denote
the ghosts for the twistor-like symmetry, scaling symmetry and worldline
supersymmetry. 

Because of the reducibility of \eqref{eq:BerkovitsTwistorLike}, $\bar{p}_{\alpha}$
is constrained to satisfy
\begin{equation}
(\lambda\gamma^{m}\gamma^{n}\bar{p})=0.
\end{equation}
However, by introducing the unconstrained spinors\begin{subequations}
\begin{eqnarray}
\theta^{\alpha} & \equiv & \bar{\theta}^{\alpha}+\gamma^{-1}\hat{\theta}^{\alpha},\\
p_{\alpha} & \equiv & \bar{p}_{\alpha}+\gamma\hat{p}_{\alpha},
\end{eqnarray}
\end{subequations}spacetime supersymmetry becomes manifest, similarly
to what was done in \cite{Berkovits:2014aia}, giving rise to the
pure spinor superparticle times a $U(1)$ decoupled sector. The partial
worldline supersymmetry can then be viewed as a tool to implement
spacetime supersymmetry.

This work is organized as follows. In section \ref{sec:linear}, the
BRST quantization of the worldline action with constraint \eqref{eq:MasslessSolution}
is analyzed and the ghost number zero cohomology is expressed as the
wave function \eqref{eq:masslessarbitraryspin}. The constraints \eqref{eq:MasslessSolution}
and \eqref{eq:BerkovitsTwistorLike} are shown to be equivalent and
the model is covariantly quantized. In section \ref{sec:PSsuperparticle},
using the Batalin-Vilkovisky formalism, the extension of Berkovits'
proposal in \cite{Berkovits:2015yra} is analyzed. The master action
is built taking into account the pure spinor constraint and derived
symmetries, which appear as constraints for the antifields. The gauge
fixing procedure leading to the action \eqref{eq:FINALPS} is carried
out and, after a simple set of field redefinitions, the pure spinor
superparticle action is obtained. Section \ref{sec:final} presents
a quick summary and some final remarks.

\section{The linear massless solution\label{sec:linear}}

After the field decomposition introduced in \eqref{eq:vectordecomposition},
the action \eqref{eq:action1st} can be simply rewritten as
\begin{equation}
S_{p}=\int_{C}d\tau\{p_{a}\dot{x}^{a}+\bar{p}^{a}\dot{\bar{x}}_{a}-e\bar{p}^{a}p_{a}\}.
\end{equation}

Now, instead of the massless constraint $\bar{p}^{a}p_{a}=0$, consider
a solution linear in the momentum of the form
\begin{equation}
\bar{p}^{a}+\gamma^{ab}p_{b}=0,\label{eq:newconstraint}
\end{equation}
where $\gamma^{ab}$ is an antisymmetric tensor. The issue with this
solution is it automatically breaks the $U(n)$ symmetry. To amend
this, $\gamma^{ab}$ is promoted to a dynamical field with conjugate
$\beta_{ab}$, such that the worldline action for this system is given
by
\begin{equation}
S=\int_{C}d\tau\{p_{a}\dot{x}^{a}+\bar{p}^{a}\dot{\bar{x}}_{a}+\tfrac{1}{2}\beta_{ab}\dot{\gamma}^{ab}-L_{a}(\bar{p}^{a}+\gamma^{ab}p_{b})\},\label{eq:newmasslessaction}
\end{equation}
where $L_{a}$ is the Lagrange multiplier of the constraint \eqref{eq:newconstraint}.
Observe that even if the term $e\bar{p}^{a}p_{a}$ is kept in the
action, a field redefinition of the form 
\begin{equation}
L_{a}\to L_{a}-ep_{a},\label{eq:einbeinshift}
\end{equation}
can absorb the \emph{einbein} $e$.

\subsection{Quantization and cohomology\label{subsec:cohomology}}

The action \eqref{eq:newmasslessaction} is invariant under the local
transformations parametrized by $\{c,c_{a},d\}$,\begin{subequations}\label{eq:bosgauge}
\begin{eqnarray}
\delta x^{a} & = & c\dot{x}^{a}-\gamma^{ab}c_{b}+d\bar{p}^{a},\\
\delta\bar{x}_{a} & = & c\dot{\bar{x}}_{a}+c_{a}+dp_{a},\\
\delta p_{a} & = & c\dot{p}_{a},\\
\delta\bar{p}^{a} & = & c\dot{\bar{p}}^{a},\\
\delta\gamma^{ab} & = & c\dot{\gamma}^{ab},\\
\delta\beta_{ab} & = & c\dot{\beta}_{ab}+(p_{a}c_{b}-p_{b}c_{a}),\\
\delta L_{a} & = & \frac{d}{d\tau}(cL_{a})+\dot{c}_{a}+\dot{d}p_{a},
\end{eqnarray}
\end{subequations}which represent respectively reparametrization
symmetry and the symmetries generated by \eqref{eq:newconstraint}
and $\bar{p}^{a}p_{a}$. However, these gauge symmetries are not all
independent, \emph{cf}. the field shift \eqref{eq:einbeinshift},
and the above transformations are invariant up to equations of motion
by\begin{subequations}\label{eq:bosgauge4gauge}
\begin{eqnarray}
\delta'c & = & \phi_{0},\\
\delta'c_{a} & = & -\phi_{0}L_{a}-\phi_{1}P_{a},\\
\delta'd & = & \phi_{1},
\end{eqnarray}
\end{subequations}with $\phi_{0}$ and $\phi_{1}$ parametrizing
the redundant gauge symmetries.

Because of the gauge-for-gauge transformations \eqref{eq:bosgauge4gauge},
the gauge symmetries parametrized by $c$ and $d$ may be disregarded.
It is straightforward then to apply the BRST construction and obtain
the gauge fixed action with its associated BRST charge. Naively, the
gauge $L_{a}=0$ could be chosen, but it is easy to see that worldline
reparametrization would reappear as a residual gauge symmetry. This
is related to the field shift \eqref{eq:einbeinshift}, for such gauge
choice would imply the degeneracy of the worldline metric. Conveniently
choosing $L_{a}=p_{a}$, the gauge fixed action is
\begin{equation}
S=\int d\tau\{p_{a}\dot{x}^{a}+\bar{p}^{a}\dot{\bar{x}}_{a}+\tfrac{1}{2}\beta_{ab}\dot{\gamma}^{ab}-\bar{p}^{a}p_{a}-b^{a}\dot{c}_{a}\},\label{eq:actiongaugefixed}
\end{equation}
with BRST charge
\begin{equation}
Q=c_{a}(\bar{p}^{a}+\gamma^{ab}p_{b}).\label{eq:BRSTcharge}
\end{equation}
Here, $c_{a}$ is the gauge parameter promoted to an anticommuting
ghost number $+1$ field and $b^{a}$ is its canonical conjugate with
ghost number $-1$.

The canonical (anti)commutation relations obtained from \eqref{eq:actiongaugefixed}
are given by\begin{subequations}
\begin{eqnarray}
[x^{a},p_{b}] & = & i\delta_{b}^{a},\\{}
[\bar{x}_{a},\bar{p}^{b}] & = & i\delta_{a}^{b},\\{}
[\gamma^{ab},\beta_{cd}] & = & i(\delta_{c}^{a}\delta_{d}^{b}-\delta_{d}^{a}\delta_{c}^{b}),\\
\{c_{a},b^{b}\} & = & i\delta_{a}^{b},
\end{eqnarray}
\end{subequations}and can be used to determine the BRST transformations
of the worldline fields,
\begin{equation}
\begin{array}{rclcrcl}
\delta x^{a} & = & c_{b}\gamma^{ba}, &  & \delta\gamma^{ab} & = & 0,\\
\delta\bar{x}_{a} & = & c_{a}, &  & \delta\beta_{ab} & = & p_{a}c_{b}-p_{b}c_{a},\\
\delta p_{a} & = & 0, &  & \delta c_{a} & = & 0,\\
\delta\bar{p}^{a} & = & 0, &  & \delta b^{a} & = & \bar{p}^{a}+\gamma^{ab}p_{b}.
\end{array}
\end{equation}

To evaluate the cohomology of $Q$, consider a generic wave function
built from $x^{a}$, $\bar{x}_{a}$, $\gamma^{ab}$ and $c_{a}$,
\begin{equation}
\Psi(c,x,\bar{x},\gamma)=F(x,\bar{x},\gamma)+c_{a}F^{a}(x,\bar{x},\gamma)+c_{a}c_{b}F^{ab}(x,\bar{x},\gamma)+\ldots
\end{equation}
where the expansion in the $c_{a}$'s is finite because of its anticommuting
character, \emph{i.e.} $(c)^{n+1}=0$. Within the BRST construction,
physical states should be annihilated by $Q$, \emph{i.e.} $Q\Psi=0$.
It is straightforward to compute the action of the BRST charge and
at ghost number zero it implies the equation of motion
\begin{equation}
\bar{\partial}^{a}F+\gamma^{ab}\partial_{b}F=0,\label{eq:eomgn0}
\end{equation}
where $\partial_{a}\equiv\frac{\partial}{\partial x^{a}}$ and $\bar{\partial}^{a}\equiv\frac{\partial}{\partial\bar{x}_{a}}$.
Other than $\bar{\partial}^{a}\partial_{a}F=0$, the above equation
does not clearly provide any dynamical information. Observe however
that by defining a new field, $\mathbb{X}^{a}$, as
\begin{equation}
\mathbb{X}^{a}\equiv x^{a}+\gamma^{ab}\bar{x}_{b},
\end{equation}
then any function $F$ of the form
\begin{equation}
F=F(\mathbb{X},\gamma),
\end{equation}
satisfies the equation of motion \eqref{eq:eomgn0}. Such solutions
were explored in \cite{Berkovits:2004bw} in generic even dimensions,
where $\gamma^{ab}$ was interpreted as a projective pure spinor.
Using the techniques introduced by Berkovits and Cherkis it is possible
to demonstrate that the wave function $F(\mathbb{X},\gamma)$ generates
massless solutions for field equations of arbitrary integer and half-integer
spin. The spin of a given solution is an extra input and appears in
the integration over the projective pure spinors.

\subsection{Projective pure spinors and the twistor-like constraint\label{subsec:projective-twistor}}

The connection between $\gamma^{ab}$ and projective pure spinors
remains to be explained. The constraint \eqref{eq:newconstraint}
can be generically expressed in $D=2n$ dimensions as a chiral spinor
constraint of the form
\begin{equation}
(\gamma^{m}\lambda)_{\alpha}P_{m}=0,\label{eq:twistor-likeconstraint}
\end{equation}
where $\gamma^{m}$ denotes a chiral block of the Dirac matrices and
$\lambda^{\alpha}$ is a projective pure spinor. The twistor-like
constraint \eqref{eq:twistor-likeconstraint} was proposed by Berkovits
in $D=10$ \cite{Berkovits:2011gh} as an attempt to explain the origin
of the pure spinor formalism \cite{Berkovits:2000fe}. 

At the level of the action, \eqref{eq:twistor-likeconstraint} is
implemented through a Lagrange multiplier $L^{\alpha}$ and it is
straightforward to show that it is equivalent to \eqref{eq:newconstraint}:
\begin{lyxlist}{00.00.0000}
\item [{$D=4$}] For $n=2$, antichiral and chiral spinors are represented
by dotted and undotted indices. In terms of $U(2)$ indices, $L_{\dot{\alpha}}$
is expressed as a vector $L_{a}$, with $a=1,2$, while a chiral spinor
is expressed as $\lambda_{\alpha}=(\lambda_{+},\lambda_{-})$. Being
a projective pure spinor, $\lambda_{\alpha}$ can be parametrized
as $\lambda_{+}=1$ and $\lambda_{-}=\gamma$. Therefore,
\begin{equation}
(L\gamma^{m}\lambda)P_{m}=L_{a}(\bar{p}^{a}+\gamma\epsilon^{ab}p_{b}).
\end{equation}
Note that in this case the antisymmetric tensor $\gamma^{ab}$ has
only one independent component, expressed here as $\gamma^{ab}\equiv\gamma\epsilon^{ab}$,
with $\epsilon^{12}=-\epsilon^{21}=1$.
\item [{$D=6$}] For $n=3$, chiral and antichiral spinors have the same
indices. The Lagrange multiplier is expressed as $L_{\alpha}=(L_{+},L_{a})$,
with $a=1,2,3$, and the projective pure spinor as $\lambda_{\alpha}=(1,\gamma_{a})$.
Therefore,
\begin{equation}
(L\gamma^{m}\lambda)P_{m}=(L_{a}+\gamma_{a}L_{+})(\bar{p}^{a}-\epsilon^{abc}p_{b}\gamma_{c}).
\end{equation}
The antisymmetric tensor $\gamma^{ab}$ can expressed as a vector
$\gamma_{a}\equiv\frac{1}{2}\epsilon_{abc}\gamma^{ab}$, where $\epsilon^{abc}$
and $\epsilon_{abc}$ are a totally antisymmetric tensors with $\epsilon^{123}=\epsilon_{321}=1$.
\item [{$D=8$}] For $n=4$, antichiral and chiral spinors are represented
by dotted and undotted indices. In terms of $U(4)$ indices, $L^{\dot{\alpha}}$
can be split into one fundamental and one antifundamental representation,
$L_{a}$ and $L^{a}$, while a projective pure spinor can be parametrized
as $\lambda^{\alpha}=(1,\gamma^{ab},\frac{1}{8}\epsilon_{abcd}\gamma^{ab}\gamma^{cd})$,
where $\epsilon_{abcd}$ is the totally antisymmetric $U(4)$ tensor
with $\epsilon_{1234}=1$. Using this parametrization, $(L\gamma^{m}\lambda)P_{m}$
can be expressed as:
\[
(L\gamma^{m}\lambda)P_{m}=(L_{a}+\tfrac{1}{2}\epsilon_{acde}\gamma^{de}L^{c})(\bar{p}^{a}+\gamma^{ab}p_{b}).
\]
\item [{$D=10$}] For $n=5$, chiral and antichiral spinors have the same
indices. The Lagrange multiplier can be expressed as $L^{\alpha}=(L^{+},L^{ab},L_{a})$,
with $a=1,\ldots,5$, and the projective pure spinor as $\lambda^{\alpha}=(1,\gamma^{ab},-\frac{1}{8}\epsilon_{abcde}\gamma^{bc}\gamma^{de})$,
in which $\epsilon_{abcde}$ is the totally antisymmetric $U(5)$
tensor with $\epsilon_{12345}=1$. Then, it is possible to show that:
\begin{equation}
(L\gamma^{m}\lambda)P_{m}=(L_{a}-\tfrac{1}{8}\epsilon_{abcde}\gamma^{bc}\gamma^{de}L^{+}+\tfrac{1}{4}\epsilon_{abcde}\gamma^{de}L^{bc})(\bar{p}^{a}+\gamma^{ab}p_{b}).\label{eq:twistorreducibilityD=00003D10}
\end{equation}
\end{lyxlist}
Notice that in every case $L_{a}$ constitutes the only independent
Lagrange multiplier, confirming that \eqref{eq:newconstraint} is
the irreducible part of the twistor-like constraint \eqref{eq:twistor-likeconstraint}.

\subsection{Covariant formulation in $D=10$\label{subsec:covariant}}

The parametrizations discussed above for $\lambda^{\alpha}$  are not Lorentz covariant. A natural way of solving this problem is to use instead a pure spinor
endowed with an extra local scaling symmetry. This solution was brought
forth in \cite{Berkovits:2015yra} through the ten-dimensional action
\begin{equation}
S_{B}=\int d\tau\{P_{m}\dot{X}^{m}+w_{\alpha}\dot{\lambda}^{\alpha}+Aw_{\alpha}\lambda^{\alpha}-(L\gamma^{m}\lambda)P_{m}\},\label{eq:Berkovitsaction}
\end{equation}
with $\alpha=1,\ldots,16$, where $\lambda^{\alpha}$ is a pure spinor
satisfying 
\begin{equation}
(\lambda\gamma^{m}\lambda)=0,\label{eq:PS}
\end{equation}
$w_{\alpha}$ is its conjugate. The Lagrange multipliers $\{L^{\alpha},A\}$
are associated respectively to the twistor-like symmetry and to a
scaling symmetry acting as
\begin{equation}
\begin{array}{rclrcl}
\delta\lambda^{\alpha} & = & \Omega\lambda^{\alpha}, & \delta w_{\alpha} & = & -\Omega w_{\alpha},\\
\delta L^{\alpha} & = & -\Omega L^{\alpha}, & \delta A & = & -\dot{\Omega},
\end{array}
\end{equation}
with parameter $\Omega$. Note that the scaling preserves the constraint
\eqref{eq:PS} and $\lambda^{\alpha}$ can then be fundamentally described
as a projective pure spinor. Furthermore, the action \eqref{eq:Berkovitsaction}
has an intrinsic gauge symmetry due to the pure spinor constraint
and it is invariant under the transformation $\delta w_{\alpha}=\epsilon_{m}(\gamma^{m}\lambda)_{\alpha}$,
with local parameter $\epsilon_{m}$. 

In \cite{Berkovits:2015yra}, using an argument based on BRST symmetry,
the reducibility of the twistor-like constraint was partially disregarded
(leading to unconstrained ghosts) and the quantization of \eqref{eq:Berkovitsaction}
was proposed to give rise to the pure spinor superparticle \cite{Berkovits:2001rb}.
However, this cannot be the case, as evidenced by the previous cohomology
analysis (subsection \ref{subsec:cohomology}). In fact, a proper
account of the gauge symmetries of the action \eqref{eq:Berkovitsaction}
uncovers a chain of symmetries connected to the pure spinor condition
\eqref{eq:PS} which, in the context of the Batalin-Vilkovisky formalism,
effectively appear as constraints on the antifields. This will be
further explored in section \ref{sec:PSsuperparticle} with the construction
of the master action in a similar setup. For now, it is enough to
know that a covariant gauge can be chosen, leading to the gauge fixed
action
\begin{equation}
S=\int d\tau\{P_{m}\dot{X}^{m}-\tfrac{1}{2}P_{m}P^{m}+w_{\alpha}\dot{\lambda}^{\alpha}+\hat{p}_{\alpha}\dot{\hat{\theta}}^{\alpha}+\bar{\Omega}\dot{\Omega}\},\label{eq:almostPSaction}
\end{equation}
where $\{\hat{\theta}^{\alpha},\hat{p}_{\alpha}\}$ is the \emph{constrained}
ghost conjugate pair for the twistor-like symmetry and $\{\Omega,\bar{\Omega}\}$
is the ghost conjugate pair for the scaling symmetry. The BRST charge
is given by 
\begin{equation}
Q=(\lambda\gamma^{m}\hat{\theta})P_{m}+\Omega(\lambda^{\alpha}w_{\alpha}+\hat{\theta}^{\alpha}\hat{p}_{\alpha}).\label{eq:almostPSBRST}
\end{equation}

The absence of spacetime supersymmetry in the action \eqref{eq:almostPSaction}
can be explained by the extra constraint connecting $\hat{p}_{\alpha}$
to $\lambda^{\alpha}$, given by
\begin{equation}
(\lambda\gamma^{m}\gamma^{n}\hat{p})=0,
\end{equation}
which arises from the reducibility of \eqref{eq:twistor-likeconstraint},
\emph{cf.} \eqref{eq:twistorreducibilityD=00003D10}, and implies
yet another intrinsic gauge symmetry for the gauge fixed action \eqref{eq:almostPSaction},\begin{subequations}
\begin{eqnarray}
\delta w_{\alpha} & = & \epsilon_{mn}(\hat{p}\gamma^{mn})_{\alpha}+\epsilon\hat{p}_{\alpha},\\
\delta\hat{\theta}^{\alpha} & = & \epsilon_{mn}(\gamma^{mn}\lambda)^{\alpha}+\epsilon\lambda^{\alpha},
\end{eqnarray}
\end{subequations}where $\epsilon$ and $\epsilon_{mn}$ are local
parameters and $\gamma^{mn}=\tfrac{1}{2}[\gamma^{m},\gamma^{n}]$.
After taking into account these symmetries, it is straightforward
to show that the BRST charge \eqref{eq:almostPSBRST} has the same
physical spectrum of the $D=10$ version of \eqref{eq:BRSTcharge}.

\section{The pure spinor superparticle\label{sec:PSsuperparticle}}

In this section a new proposal is presented to extend Berkovits' twistor-like
constraint with (partial) worldline supersymmetry, finally leading
to the pure spinor superparticle.

Consider an extension of \eqref{eq:Berkovitsaction} with the inclusion
of an anticommuting variable $\bar{\theta}^{\alpha}$ satisfying 
\begin{equation}
(\lambda\gamma^{m}\bar{\theta})=0,\label{eq:SUSYPS}
\end{equation}
together with a sort of worldline supersymmetry for the spacetime
spinors. Such a model can be expressed by the action
\begin{equation}
S_{0}=\int d\tau\{P_{m}\dot{X}^{m}+w_{\alpha}\nabla\lambda^{\alpha}-L^{\alpha}(\gamma^{m}\lambda)_{\alpha}P_{m}+\bar{p}_{\alpha}\dot{\bar{\theta}}^{\alpha}+\chi\lambda^{\alpha}\bar{p}_{\alpha}\},\label{eq:PSunfixed}
\end{equation}
where $\bar{p}_{\alpha}$ is the conjugate of $\bar{\theta}^{\alpha}$,
$\chi$ is the Lagrange multiplier of the local symmetry generator
$\lambda^{\alpha}\bar{p}_{\alpha}$ and $\nabla$ is the covariant
derivative for the scaling symmetry with gauge field $A$.

Due to the constraints \eqref{eq:PS} and \eqref{eq:SUSYPS}, the
action is invariant under the transformations\begin{subequations}\label{eq:PSsymmetries}
\begin{eqnarray}
\delta w_{\alpha} & = & d_{m}(\gamma^{m}\lambda)_{\alpha}+e_{m}(\gamma^{m}\bar{\theta})_{\alpha},\\
\delta\bar{p}_{\alpha} & = & e_{m}(\gamma^{m}\lambda)_{\alpha},\\
\delta L^{\alpha} & = & f\lambda^{\alpha}+f_{mn}(\gamma^{mn}\lambda)^{\alpha}+g\bar{\theta}^{\alpha},
\end{eqnarray}
\end{subequations}where $d_{m}$, $e_{m}$, $f$ , $f_{mn}$ and
$g$ are local parameters. From now on, these will be called pure
spinor symmetries. Having a special role, they are not to be gauge
fixed and instead appear as constraints on the antifields in the construction
of the master action.

Keeping this in mind, the gauge symmetries of \eqref{eq:PSunfixed}
are summarized by\begin{subequations}\label{eq:PSgauge}
\begin{eqnarray}
\delta X^{m} & = & c\dot{X}^{m}+(\lambda\gamma^{m}\hat{\theta}),\\
\delta P_{m} & = & c\dot{P}_{m},\\
\delta\lambda^{\alpha} & = & c\dot{\lambda}^{\alpha}+\Omega\lambda^{\alpha},\\
\delta w_{\alpha} & = & c\dot{w}_{\alpha}-P_{m}(\gamma^{m}\hat{\theta})_{\alpha}-\Omega w_{\alpha}+\gamma\bar{p}_{\alpha},\\
\delta L^{\alpha} & = & \frac{d}{d\tau}(cL^{\alpha})+\nabla\hat{\theta}^{\alpha}-\Omega L^{\alpha},\\
\delta A & = & \frac{d}{d\tau}(cA)-\dot{\Omega},\\
\delta\bar{\theta}^{\alpha} & = & c\dot{\bar{\theta}}^{\alpha}+\gamma\lambda^{\alpha},\\
\delta\bar{p}_{\alpha} & = & c\dot{\bar{p}}_{\alpha},\\
\delta\chi & = & \frac{d}{d\tau}(c\chi)+\nabla\gamma-\Omega\chi,
\end{eqnarray}
\end{subequations}where the parameters $c$, $\Omega$, $\hat{\theta}^{\alpha}$
and $\gamma$ respectively denote worldline reparametrization, scaling
symmetry, twistor-like symmetry and the worldline supersymmetry of
the spacetime spinors. These gauge symmetries, however, are not irreducible.
Consider the following transformations of the gauge parameters,\begin{subequations}\label{eq:PSgauge-for-gauge}
\begin{eqnarray}
\delta'c & = & \phi,\\
\delta'\Omega & = & \phi A,\\
\delta'\hat{\theta}^{\alpha} & = & -\phi L^{\alpha},\\
\delta'\gamma & = & -\phi\chi.
\end{eqnarray}
\end{subequations}Here, $\phi$ is the gauge-for-gauge parameter.
It is straightforward to show that they leave the gauge transformations
invariant up to equations of motion,
\begin{equation}
\begin{array}{rclcrcl}
\delta'[\delta X^{m}] & = & \phi\{\dot{X}^{m}-(\lambda\gamma^{m}L)\}, &  & \delta'[\delta A] & = & 0,\\
\delta'[\delta P_{m}] & = & \phi\{\dot{P}_{m}\}, &  & \delta'[\delta\bar{\theta}^{\alpha}] & = & \phi\{\nabla\bar{\theta}^{\alpha}-\chi\lambda^{\alpha}\}\\
\delta'[\delta\lambda^{\alpha}] & = & \phi\{\nabla\lambda^{\alpha}\}, &  & \delta'[\delta\bar{p}_{\alpha}] & = & \phi\{\nabla\bar{p}_{\alpha}\},\\
\delta'[\delta w_{\alpha}] & = & \phi\{\nabla w_{\alpha}-\chi\bar{p}_{\alpha}+P_{m}(\gamma^{m}L)_{\alpha}\}, &  & \delta'[\delta\chi] & = & 0.
\end{array}\label{eq:PSgauge-for-gaugeEOM}
\end{equation}

\subsection{Master action}

The Batalin-Vilkovisky (BV) formalism seems to be the most suitable
tool to deal with the gauge structure of the action \eqref{eq:PSunfixed}.
In order to build the master action, the field content of the model
is extended by promoting gauge and gauge-for-gauge parameters to fields.
They will be collectively denoted by $\Phi^{I}$, where $I$ is an
index running over all fields:
\begin{equation}
\Phi^{I}\equiv\{X^{m},P_{m},\lambda^{\alpha},w_{\alpha},L^{\alpha},A,\bar{\theta}^{\alpha},\bar{p}_{\alpha},\chi,c,\Omega,\hat{\theta}^{\alpha},\gamma,\phi\}.
\end{equation}
Note that $\{c,\Omega,\hat{\theta}^{\alpha}\}$ are Grassmann odd
while $\{\gamma,\phi\}$ are Grassmann even fields. The antifields,
denoted by $\Phi_{I}^{*}$, have the opposite statistics of the fields
and will be defined as
\begin{equation}
\Phi_{I}^{*}\equiv\{X_{m}^{*},P_{*}^{m},\lambda_{\alpha}^{*},w_{*}^{\alpha},L_{\alpha}^{*},A^{*},\bar{\theta}_{\alpha}^{*},\bar{p}_{*}^{\alpha},\chi^{*},c^{*},\Omega^{*},\hat{\theta}_{\alpha}^{*},\gamma^{*},\phi^{*}\}.
\end{equation}

Fields and antifields are used to define the antibrackets between
any operator pair $\mathcal{O}_{1}$ and $\mathcal{O}_{2}$,
\begin{equation}
\{\mathcal{O}_{1},\mathcal{O}_{2}\}\equiv\mathcal{O}_{1}\left(\frac{\overleftarrow{\partial}}{\partial\Phi_{I}^{*}}\frac{\partial}{\partial\Phi^{I}}-\frac{\overleftarrow{\partial}}{\partial\Phi^{I}}\frac{\partial}{\partial\Phi_{I}^{*}}\right)\mathcal{O}_{2},
\end{equation}
where the sum over $I$ is implicit. In particular,
\begin{equation}
\{\Phi_{I}^{*},\Phi^{J}\}=\delta_{I}^{J}.\label{eq:FAFconjugation}
\end{equation}

Due to the pure spinor symmetries of \eqref{eq:PSsymmetries}, the
antifields are constrained to satisfy\begin{subequations}\label{eq:PSantifieldconstraints}
\begin{eqnarray}
(\lambda\gamma^{m}w_{*}) & = & 0,\\
(\lambda\gamma^{m}\bar{p}_{*})+(\bar{\theta}\gamma^{m}w_{*}) & = & 0,\\
\bar{\theta}^{\alpha}L_{\alpha}^{*} & = & 0,\\
(\lambda\gamma^{m}\gamma^{n}L^{*}) & = & 0.
\end{eqnarray}
\end{subequations}Observe that treating \eqref{eq:PSsymmetries}
as ordinary gauge transformations leads to an infinite tower of gauge-for-gauge
symmetries. Instead, the problem is simplified if the above conditions
are imposed. There is an apparent contradiction between the constraints
\eqref{eq:PS}, \eqref{eq:SUSYPS} and \eqref{eq:PSantifieldconstraints},
and the field-antifield conjugation operation \eqref{eq:FAFconjugation}.
For example,
\begin{eqnarray}
\{\lambda_{\alpha}^{*},(\lambda\gamma^{m}\lambda)\} & = & 2(\gamma^{m}\lambda)_{\alpha},\nonumber \\
 & \neq & 0,
\end{eqnarray}
but the pure spinor (gauge) symmetries implied by the constraints
have to be taken into account. This is a known feature of the pure
spinor beta-gamma system and it can be easily solved by using only
gauge invariant operators. The pure spinor symmetries will be further
discussed later.

The master action can be cast as
\begin{equation}
S=S_{0}+S_{1}+S_{2}+S_{3},\label{eq:PSmaster}
\end{equation}
where $S_{0}$ is displayed in \eqref{eq:PSunfixed} and\begin{subequations}
\begin{eqnarray}
S_{1} & = & \int d\tau\{c\dot{X}^{m}X_{m}^{*}+c\dot{P}_{m}P_{*}^{m}+c\dot{\lambda}^{\alpha}\lambda_{\alpha}^{*}+c\dot{w}_{\alpha}w_{*}^{\alpha}-cL^{\alpha}\dot{L}_{\alpha}^{*}-cA\dot{A}^{*}\}\nonumber \\
 &  & +\int d\tau\{c\dot{\bar{p}}_{\alpha}\bar{p}_{*}^{\alpha}+c\dot{\bar{\theta}}^{\alpha}\bar{\theta}_{\alpha}^{*}-c\chi\dot{\chi}^{*}+\gamma\lambda^{\alpha}\bar{\theta}_{\alpha}^{*}+\gamma\bar{p}_{\alpha}w_{*}^{\alpha}+(\nabla\gamma)\chi^{*}\}\nonumber \\
 &  & +\int d\tau\{(\lambda\gamma^{m}\hat{\theta})X_{m}^{*}-P_{m}(\hat{\theta}\gamma^{m}w_{*})+\nabla\hat{\theta}^{\alpha}L_{\alpha}^{*}\}\nonumber \\
 &  & +\int d\tau\{\Omega\lambda^{\alpha}\lambda_{\alpha}^{*}-\Omega w_{\alpha}w_{*}^{\alpha}-\Omega L^{\alpha}L_{\alpha}^{*}-\Omega\chi\chi^{*}-\dot{\Omega}A^{*}\},\\
S_{2} & = & \int d\tau\{c\dot{c}c^{*}+c\dot{\Omega}\Omega^{*}+c\dot{\gamma}\gamma^{*}+c\dot{\hat{\theta}}^{\alpha}\hat{\theta}_{\alpha}^{*}-\Omega\hat{\theta}^{\alpha}\hat{\theta}_{\alpha}^{*}-\Omega\gamma\gamma^{*}\}\nonumber \\
 &  & +\int d\tau\{\phi c^{*}+\phi A\Omega^{*}-\phi L^{\alpha}\hat{\theta}_{\alpha}^{*}-\phi\chi\gamma^{*}+c\dot{\phi}\phi^{*}-\dot{c}\phi\phi^{*}\},\\
S_{3} & = & \int d\tau\{\phi P_{*}^{m}X_{m}^{*}+\phi w_{*}^{\alpha}\lambda_{\alpha}^{*}-\phi\bar{\theta}_{\alpha}^{*}\bar{p}_{*}^{\alpha}\}.
\end{eqnarray}
\end{subequations}$S_{1}$ accounts for the gauge transformations
\eqref{eq:PSgauge}, $S_{2}$ is associated to the extension of the
gauge algebra to the ghost fields and the gauge-for-gauge symmetries
\eqref{eq:PSgauge-for-gauge}, and $S_{3}$ (quadratic in the antifields)
is necessary because of the on-shell invariance of the gauge transformations,
\emph{cf.} \eqref{eq:PSgauge-for-gaugeEOM}.

The master action \eqref{eq:PSmaster} satisfies the master equation,
\begin{equation}
\{S,S\}=0,
\end{equation}
and, for any operator $\mathcal{O}$, it defines its BV transformation,
given by
\begin{equation}
\delta_{\text{\tiny{BV}}}\mathcal{O}\equiv\{S,\mathcal{O}\}.
\end{equation}
Satisfying the master equation is equivalent to the statement that
the master action is invariant under the BV transformations. For completeness,
the transformations of the fields are\begin{subequations}\label{eq:BVfields}
\begin{eqnarray}
\delta_{\text{\tiny{BV}}}X^{m} & = & c\dot{X}^{m}+(\lambda\gamma^{m}\hat{\theta})+\phi P_{*}^{m},\\
\delta_{\text{\tiny{BV}}}P_{m} & = & c\dot{P}_{m}-\phi X_{m}^{*},\\
\delta_{\text{\tiny{BV}}}\lambda^{\alpha} & = & c\dot{\lambda}^{\alpha}+\Omega\lambda^{\alpha}+\phi w_{*}^{\alpha},\\
\delta_{\text{\tiny{BV}}}w_{\alpha} & = & c\dot{w}_{\alpha}-P_{m}(\hat{\theta}\gamma^{m})_{\alpha}-\Omega w_{\alpha}+\gamma\bar{p}_{\alpha}-\phi\lambda_{\alpha}^{*},\\
\delta_{\text{\tiny{BV}}}L^{\alpha} & = & \frac{d}{d\tau}(cL^{\alpha})+\nabla\hat{\theta}^{\alpha}-\Omega L^{\alpha},\\
\delta_{\text{\tiny{BV}}}A & = & \frac{d}{d\tau}(cA)-\dot{\Omega},\\
\delta_{\text{\tiny{BV}}}\bar{\theta}^{\alpha} & = & c\dot{\bar{\theta}}^{\alpha}+\gamma\lambda^{\alpha}-\phi\bar{p}_{*}^{\alpha},\\
\delta_{\text{\tiny{BV}}}\bar{p}_{\alpha} & = & c\dot{\bar{p}}_{\alpha}-\phi\bar{\theta}_{\alpha}^{*},\\
\delta_{\text{\tiny{BV}}}\chi & = & \frac{d}{d\tau}(c\chi)+\nabla\gamma-\Omega\chi,\\
\delta_{\text{\tiny{BV}}}c & = & c\dot{c}+\phi,\\
\delta_{\text{\tiny{BV}}}\Omega & = & c\dot{\Omega}+\phi A,\\
\delta_{\text{\tiny{BV}}}\hat{\theta}^{\alpha} & = & c\dot{\hat{\theta}}^{\alpha}-\Omega\hat{\theta}^{\alpha}-\phi L^{\alpha},\\
\delta_{\text{\tiny{BV}}}\gamma & = & c\dot{\gamma}-\Omega\gamma-\phi\chi,\\
\delta_{\text{\tiny{BV}}}\phi & = & c\dot{\phi}-\dot{c}\phi,
\end{eqnarray}
\end{subequations}and the transformation of the antifields are\begin{subequations}\label{eq:BVantifields}
\begin{eqnarray}
\delta_{\text{\tiny{BV}}}X_{m}^{*} & = & \dot{P}_{m}+\frac{d}{d\tau}(cX_{m}^{*}),\\
\delta_{\text{\tiny{BV}}}P_{*}^{m} & = & -\dot{X}^{m}+(L\gamma^{m}\lambda)+\frac{d}{d\tau}(cP_{*}^{m})+(\hat{\theta}\gamma^{m}w_{*}),\\
\delta_{\text{\tiny{BV}}}\lambda_{\alpha}^{*} & = & \nabla w_{\alpha}+(\gamma^{m}L)_{\alpha}P_{m}-\chi\bar{p}_{\alpha}+\frac{d}{d\tau}(c\lambda_{\alpha}^{*})-(\gamma^{m}\hat{\theta})_{\alpha}X_{m}^{*}-\gamma\bar{\theta}_{\alpha}^{*}-\Omega\lambda_{\alpha}^{*},\\
\delta_{\text{\tiny{BV}}}w_{*}^{\alpha} & = & -\nabla\lambda^{\alpha}+\frac{d}{d\tau}(cw_{*}^{\alpha})+\Omega w_{*}^{\alpha},\\
\delta_{\text{\tiny{BV}}}L_{\alpha}^{*} & = & (\gamma^{m}\lambda)_{\alpha}P_{m}+c\dot{L}_{\alpha}^{*}+\Omega L_{\alpha}^{*}+\phi\hat{\theta}_{\alpha}^{*},\\
\delta_{\text{\tiny{BV}}}A^{*} & = & c\dot{A}^{*}-w_{\alpha}\lambda^{\alpha}+\hat{\theta}^{\alpha}L_{\alpha}^{*}+\gamma\chi^{*}-\phi\Omega^{*},\\
\delta_{\text{\tiny{BV}}}\bar{\theta}_{\alpha}^{*} & = & \dot{\bar{p}}_{\alpha}+\frac{d}{d\tau}(c\bar{\theta}_{\alpha}^{*}),\\
\delta_{\text{\tiny{BV}}}\bar{p}_{*}^{\alpha} & = & \dot{\bar{\theta}}^{\alpha}-\chi\lambda^{\alpha}+\frac{d}{d\tau}(c\bar{p}_{*}^{\alpha})+\gamma w_{*}^{\alpha},\\
\delta_{\text{\tiny{BV}}}\chi^{*} & = & \lambda^{\alpha}\bar{p}_{\alpha}+c\dot{\chi}^{*}+\Omega\chi^{*}-\phi\gamma^{*},\\
\delta_{\text{\tiny{BV}}}c^{*} & = & \dot{X}^{m}X_{m}^{*}+\dot{P}_{m}P_{m}^{*}+\dot{\lambda}^{\alpha}\lambda_{\alpha}^{*}+\dot{w}_{\alpha}w_{*}^{\alpha}-L^{\alpha}\dot{L}_{\alpha}^{*}-A\dot{A}^{*}+\dot{\bar{p}}_{\alpha}\bar{p}_{*}^{\alpha}+\dot{\bar{\theta}}^{\alpha}\bar{\theta}_{\alpha}^{*}\nonumber \\
 &  & -\chi\dot{\chi}^{*}+2\dot{c}c^{*}+c\dot{c}^{*}+\dot{\gamma}\gamma^{*}+\dot{\hat{\theta}}^{\alpha}\hat{\theta}_{\alpha}^{*}+\dot{\Omega}\Omega^{*}+2\dot{\phi}\phi^{*}+\phi\dot{\phi}^{*},\\
\delta_{\text{\tiny{BV}}}\Omega^{*} & = & \lambda^{\alpha}\lambda_{\alpha}^{*}-w_{\alpha}w_{*}^{\alpha}-L^{\alpha}L_{\alpha}^{*}+\dot{A}^{*}-\chi\chi^{*}+\frac{d}{d\tau}(c\Omega^{*})-\hat{\theta}^{\alpha}\hat{\theta}_{\alpha}^{*}-\gamma\gamma^{*},\\
\delta_{\text{\tiny{BV}}}\hat{\theta}_{\alpha}^{*} & = & (\lambda\gamma^{m})_{\alpha}X_{m}^{*}-P_{m}(\gamma^{m}w_{*})_{\alpha}-\nabla L_{\alpha}^{*}+\frac{d}{d\tau}(c\hat{\theta}_{\alpha}^{*})+\Omega\hat{\theta}_{\alpha}^{*},\\
\delta_{\text{\tiny{BV}}}\gamma^{*} & = & \nabla\chi^{*}-\lambda^{\alpha}\bar{\theta}_{\alpha}^{*}-\bar{p}_{\alpha}w_{*}^{\alpha}+\frac{d}{d\tau}(c\gamma^{*})+\Omega\gamma^{*},\\
\delta_{\text{\tiny{BV}}}\phi^{*} & = & 2\dot{c}\phi^{*}+c\dot{\phi}^{*}-c^{*}-A\Omega^{*}+L^{\alpha}\hat{\theta}_{\alpha}^{*}+\chi\gamma^{*}+\bar{\theta}_{\alpha}^{*}\bar{p}_{*}^{\alpha}-P_{*}^{m}X_{m}^{*}-w_{*}^{\alpha}\lambda_{\alpha}^{*}.
\end{eqnarray}
\end{subequations}

As mentioned before, the pure spinor conditions \eqref{eq:PS} and
\eqref{eq:SUSYPS} imply a series of additional constraints on the
antifields which can be summarized as\begin{subequations}\label{eq:PSconstraints}
\begin{eqnarray}
(\lambda\gamma^{m}w_{*}) & = & 0,\\
(\lambda\gamma^{m}\bar{p}_{*})+(\bar{\theta}\gamma^{m}w_{*}) & = & 0,\\
\bar{\theta}^{\alpha}L_{\alpha}^{*} & = & 0,\\
\lambda^{\alpha}L_{\alpha}^{*} & = & 0,\\
(\lambda\gamma^{mn}L^{*}) & = & 0,\\
\bar{p}_{*}^{\alpha}L_{\alpha}^{*}+\bar{\theta}^{\alpha}\hat{\theta}_{\alpha}^{*} & = & 0,\\
\lambda^{\alpha}\hat{\theta}_{\alpha}^{*}+w_{*}^{\alpha}L_{\alpha}^{*} & = & 0,\\
(\lambda\gamma^{mn}\hat{\theta}^{*})+(w_{*}\gamma^{mn}L^{*}) & = & 0.
\end{eqnarray}
\end{subequations}It is easy to check their consistency with the
BV transformations. Naturally, such constraints are associated to
the pure spinor gauge transformations and the master action is invariant
under the pure spinor symmetries\begin{subequations}\label{eq:PSsymmetriesFULL}
\begin{eqnarray}
\delta w_{\alpha} & = & d_{m}(\gamma^{m}\lambda)_{\alpha}+e_{m}(\gamma^{m}\bar{\theta})_{\alpha}-\bar{f}L_{\alpha}^{*}+\bar{f}_{mn}(\gamma^{mn}L^{*})_{\alpha},\\
\delta\bar{p}_{\alpha} & = & e_{m}(\gamma^{m}\lambda)_{\alpha}+\bar{g}L_{\alpha}^{*},\\
\delta L^{\alpha} & = & f\lambda^{\alpha}+f_{mn}(\gamma^{mn}\lambda)^{\alpha}+g\bar{\theta}^{\alpha}+\bar{f}w_{*}^{\alpha}+\bar{f}_{mn}(\gamma^{mn}w_{*})^{\alpha}+\bar{g}\bar{p}_{*}^{\alpha},\\
\delta\hat{\theta}^{\alpha} & = & \bar{f}\lambda^{\alpha}+\bar{f}_{mn}(\gamma^{mn}\lambda)^{\alpha}+\bar{g}\bar{\theta}^{\alpha},\\
\delta\lambda_{\alpha}^{*} & = & b_{m}(\gamma^{m}\lambda)_{\alpha}+c_{m}(\gamma^{m}\bar{\theta})_{\alpha}-d_{m}(\gamma^{m}w_{*})_{\alpha}-e_{m}(\gamma^{m}\bar{p}_{*})_{\alpha}\nonumber \\
 &  & -fL_{\alpha}^{*}+f_{mn}(\gamma^{mn}L^{*})_{\alpha}-\bar{f}\hat{\theta}_{\alpha}^{*}+\bar{f}_{mn}(\gamma^{mn}\hat{\theta}^{*})_{\alpha},\\
\delta\bar{\theta}_{\alpha}^{*} & = & c_{m}(\gamma^{m}\lambda)_{\alpha}+e_{m}(\gamma^{m}w_{*})_{\alpha}+gL_{\alpha}^{*}-\bar{g}\hat{\theta}_{\alpha}^{*},
\end{eqnarray}
\end{subequations}where $b_{m}$, $c_{m}$, $d_{m}$, $e_{m}$, $f$,
$f_{mn}$, $g$, $\bar{f}$ , $\bar{f}_{mn}$ and $\bar{g}$ are local
parameters.

\subsection{Gauge fixing}

A systematic procedure to fix the gauge symmetries of the theory involves
the introduction of the non-minimal master action through auxiliary
variables organized as antighosts, $\bar{\Phi}_{i}$, Nakanishi-Lautrup
fields, $\Lambda_{i}$, and their respective antifields, $\bar{\Phi}_{*}^{i}$
and $\Lambda_{*}^{i}$, where the index $i$ usually denotes the gauge
and gauge-for-gauge symmetries to be fixed. The non-minimal action
is defined as
\begin{equation}
\tilde{S}=S+\int d\tau\,\bar{\Phi}_{*}^{i}\Lambda_{i},
\end{equation}
and, by construction, it satisfies the master equation in the extended
phase space.

In the present case,\begin{subequations}
\begin{eqnarray}
\bar{\Phi}_{i} & \equiv & \{b,\bar{\Omega},\hat{p}_{\alpha},\beta,\bar{\phi}\},\\
\Lambda_{i} & \equiv & \{B,O,M_{\alpha},G,F\},\\
\bar{\Phi}_{*}^{i} & \equiv & \{b^{*},\bar{\Omega}^{*},\hat{p}_{*}^{\alpha},\beta^{*},\bar{\phi}^{*}\},\\
\Lambda_{*}^{i} & \equiv & \{B^{*},O^{*},M_{*}^{\alpha},G^{*},F^{*}\},
\end{eqnarray}
\end{subequations}where $\{b,\bar{\Omega},\hat{p}_{\alpha},\beta,\bar{\phi}\}$
are respectively the antighosts of $\{c,\Omega,\hat{\theta}^{\alpha},\gamma,\phi\}$,
and so on.

The gauge fixing fermion will be chosen to be
\begin{equation}
\Xi=\int d\tau\left\{ \hat{p}_{\alpha}\left[P_{m}\frac{(\gamma^{m}C)^{\alpha}}{2(C\lambda)}-L^{\alpha}\right]+\bar{\phi}c+\beta\chi+\bar{\Omega}A\right\} ,\label{eq:gaugefixingfermion}
\end{equation}
imposing the gauges $A=c=\chi=0$ and
\begin{equation}
L^{\alpha}=P_{m}\frac{(\gamma^{m}C)^{\alpha}}{2(C\lambda)},
\end{equation}
where $C_{\alpha}$ is a constant spinor. The particular choice for
$L^{\alpha}$ is related to the discussion just before equation \eqref{eq:actiongaugefixed}
and the degeneracy of the worldline metric, and the gauge-for-gauge
symmetry is being used to eliminate the reparametrization symmetry.
At the level of the non-minimal master action, the gauge fixing is
implemented by evaluating $\tilde{S}$ at\begin{subequations}
\begin{eqnarray}
\Phi_{I}^{*} & = & \frac{\delta\Xi}{\delta\Phi^{I}},\\
\bar{\Phi}_{*}^{i} & = & \frac{\delta\Xi}{\delta\Phi_{i}},\\
\Lambda_{*}^{i} & = & \frac{\delta\Xi}{\delta\Lambda_{i}},
\end{eqnarray}
\end{subequations}and the non-vanishing antifields are given here
by\begin{subequations}
\begin{eqnarray}
P_{*}^{m} & = & \frac{(\hat{p}\gamma^{m}C)}{2(C\lambda)},\\
\lambda_{\alpha}^{*} & = & -P_{m}\frac{(\hat{p}\gamma^{m}C)}{2(C\lambda)^{2}}C_{\alpha},\\
L_{\alpha}^{*} & = & -\hat{p}_{\alpha}\\
A^{*} & = & \bar{\Omega},\\
\chi^{*} & = & \beta,\\
c^{*} & = & \bar{\phi},\\
\bar{\Omega}^{*} & = & A,\\
\beta^{*} & = & \chi,\\
\bar{\phi}^{*} & = & c.
\end{eqnarray}
\end{subequations}Note that these solutions have to be consistent
with the antifield constraints of \eqref{eq:PSconstraints}, thus\begin{subequations}
\begin{eqnarray}
\bar{\theta}^{\alpha}\hat{p}_{\alpha} & = & 0,\\
\lambda^{\alpha}\hat{p}_{\alpha} & = & 0,\\
(\lambda\gamma^{mn}\hat{p}) & = & 0.
\end{eqnarray}
\end{subequations}

The gauge-fixed action, after solving for the equations of motion
of the Nakanishi-Lautrup fields, is simply
\begin{equation}
S_{\tiny{PS}}=\int d\tau\{P_{m}\dot{X}^{m}+w_{\alpha}\dot{\lambda}^{\alpha}-\frac{1}{2}P_{m}P^{m}+\bar{p}_{\alpha}\dot{\bar{\theta}}^{\alpha}+\hat{p}_{\alpha}\dot{\hat{\theta}}^{\alpha}+\bar{\Omega}\dot{\Omega}+\beta\dot{\gamma}\},\label{eq:PSfinal-scale}
\end{equation}
and it is invariant under the pure spinor symmetries
\begin{eqnarray}
\delta w_{\alpha} & = & d_{m}(\gamma^{m}\lambda)_{\alpha}+e_{m}(\gamma^{m}\bar{\theta})_{\alpha}+\bar{f}\hat{p}_{\alpha}-\bar{f}_{mn}(\gamma^{mn}\hat{p})_{\alpha},\\
\delta\bar{p}_{\alpha} & = & e_{m}(\gamma^{m}\lambda)_{\alpha}-\bar{g}\hat{p}_{\alpha},\\
\delta\hat{\theta}^{\alpha} & = & \bar{f}\lambda^{\alpha}+\bar{f}_{mn}(\gamma^{mn}\lambda)^{\alpha}+\bar{g}\bar{\theta}^{\alpha}.
\end{eqnarray}

The BV-BRST transformations can be easily read from \eqref{eq:BVfields}
and \eqref{eq:BVantifields} and are given by\begin{subequations}
\begin{eqnarray}
\delta X^{m} & = & (\lambda\gamma^{m}\hat{\theta}),\\
\delta P_{m} & = & 0,\\
\delta\lambda^{\alpha} & = & \Omega\lambda^{\alpha},\\
\delta w_{\alpha} & = & -P_{m}(\hat{\theta}\gamma^{m})_{\alpha}-\Omega w_{\alpha}+\gamma\bar{p}_{\alpha},\\
\delta\bar{\theta}^{\alpha} & = & \gamma\lambda^{\alpha},\\
\delta\bar{p}_{\alpha} & = & 0,\\
\delta\Omega & = & 0,\\
\delta\bar{\Omega} & = & -w_{\alpha}\lambda^{\alpha}+\beta\gamma+\hat{p}_{\alpha}\hat{\theta}^{\alpha},\\
\delta\hat{\theta}^{\alpha} & = & -\Omega\hat{\theta}^{\alpha},\\
\delta\hat{p}_{\alpha} & = & -(\gamma^{m}\lambda)_{\alpha}P_{m}+\Omega\hat{p}_{\alpha},\\
\delta\gamma & = & -\Omega\gamma,\\
\delta\beta & = & \lambda^{\alpha}\bar{p}_{\alpha}+\Omega\beta.
\end{eqnarray}
\end{subequations}Using Noether's theorem, the BRST charge is computed
to be
\begin{equation}
Q=\gamma\lambda^{\alpha}\bar{p}_{\alpha}-(\lambda\gamma^{m}\hat{\theta})P_{m}+\Omega(\beta\gamma-\hat{\theta}^{\alpha}\hat{p}_{\alpha}-\lambda^{\alpha}w_{\alpha}).\label{eq:BRST-PS-untwisted}
\end{equation}
Note that the terms involving the constant spinor $C_{\alpha}$ decouple
from the action and from the transformations above, as they are proportional
to the on-shell vanishing ghost-for-ghost $\phi$.

\subsection{Ghost number twisting and cohomology}

The gauge fixed action can be expressed in terms of an unconstrained
spacetime spinor $\theta^{\alpha}$, making spacetime supersymmetry
manifest.

To see this, consider the field redefinitions\begin{subequations}\label{eq:fieldredefinition}
\begin{eqnarray}
\lambda^{\alpha} & \to & \gamma^{-1}\lambda^{\alpha},\\
w_{\alpha} & \to & \gamma w_{\alpha},\\
\hat{\theta}^{\alpha} & \to & \gamma\hat{\theta}^{\alpha},\\
\hat{p}_{\alpha} & \to & \gamma^{-1}\hat{p}_{\alpha},\\
\beta & \to & \beta+\gamma^{-1}\lambda^{\alpha}w_{\alpha}+\gamma^{-1}\hat{\theta}^{\alpha}\hat{p}_{\alpha}.
\end{eqnarray}
\end{subequations}Although the action \eqref{eq:PSfinal-scale} is
left unchanged, all spinors are now invariant under the scale transformations
($\Omega$) with a consequent shift of their ghost number: $\lambda^{\alpha}$($w_{\alpha}$)
has now ghost number $1$ ($-1$), while $\hat{\theta}^{\alpha}$,
$\hat{p}_{\alpha}$, $\bar{\theta}^{\alpha}$, and $\bar{p}_{\alpha}$
have ghost number $0$.

The spacetime spinors $\{\hat{\theta}^{\alpha},\hat{p}_{\alpha},\bar{\theta}^{\alpha},\bar{p}_{\alpha}\}$
can then combine into an unconstrained conjugate pair $\{\theta^{\alpha},p_{\alpha}\}$,
with the emergence of spacetime supersymmetry. The unconstrained spinors
are defined as\begin{subequations}
\begin{eqnarray}
\theta^{\alpha} & \equiv & \hat{\theta}^{\alpha}+\bar{\theta}^{\alpha},\\
p_{\alpha} & \equiv & \hat{p}_{\alpha}+\bar{p}_{\alpha}.
\end{eqnarray}
\end{subequations}To show that $\{\theta^{\alpha},p_{\alpha}\}$
are indeed unconstrained, first recall that, by construction, $\bar{p}_{\alpha}$
has a gauge invariance of the form $\delta\bar{p}_{\alpha}=e_{m}(\gamma^{m}\lambda)_{\alpha}$
due to the constraint $(\lambda\gamma^{m}\bar{\theta})=0$. This leaves
$\bar{\theta}^{\alpha}$ and $\bar{p}_{\alpha}$ with 11 independent
components each. Due to the constraint $(\lambda\gamma^{mn}\hat{p})=0$,
$\hat{p}_{\alpha}$ has only 5 independent components, while the number
of irreducible components in the twistor-like constraint (and consequently
the number of independent components of the associated ghost, $\hat{\theta}^{\alpha}$)
is also 5, \emph{cf}. section \ref{sec:linear}. Their complementarity
($11+5=16$) can be directly shown by the $SO(10)$ spinor decomposition.

With the redefinition \eqref{eq:fieldredefinition}, note also that
the ghost variables $\{\gamma,\beta,\Omega,\bar{\Omega}\}$ constitute
a decoupled $U(1)$ sector. After the addition of the null terms $\lambda^{\alpha}\hat{p}_{\alpha}$
and $(\lambda\gamma^{m}\bar{\theta})P_{m}$, the BRST charge can be
rewritten as
\begin{equation}
Q=Q_{*}+Q_{\tiny{PS}},\label{eq:QPS+U1}
\end{equation}
where\begin{subequations}
\begin{eqnarray}
Q_{*} & = & \Omega\beta\gamma,\\
Q_{\tiny{PS}} & = & \lambda^{\alpha}d_{\alpha}.\label{eq:QPS}
\end{eqnarray}
\end{subequations}$Q_{\tiny{PS}}$ is the pure spinor BRST charge,with
\begin{equation}
d_{\alpha}\equiv p_{\alpha}-P_{m}(\gamma^{m}\theta)_{\alpha}.
\end{equation}

Since $Q_{\tiny{PS}}$ and $Q_{*}$ are independent of each other,
it is more convenient to analyze their cohomology separately. There
are two elements in the cohomology of $Q_{*}$, the identity operator,
$\mathbbm{1}$, and $\Omega$. Any other $Q_{*}$-closed operator
can be shown to be $Q_{*}$-exact. In fact, even the operator $\Omega$
is singular in the sense that it can be written as the limit of a
$Q_{*}$-exact operator:
\begin{equation}
\lim_{\epsilon\to0}\left\{ Q_{*},\frac{1}{\epsilon}\gamma^{\epsilon}\right\} =\Omega.
\end{equation}

The cohomology of $Q_{\tiny{PS}}$ is well known \cite{Berkovits:2001rb}
and will be just quickly reviewed here. According to the pure spinor
ghost number, it is organized as
\begin{equation}
\left\{ \begin{array}{cccc}
\mathbbm{1}, & \lambda^{\alpha}A_{\alpha}, & \lambda^{\alpha}\lambda^{\beta}A_{\alpha\beta}, & (\lambda\gamma^{m}\theta)(\lambda\gamma^{n}\theta)(\lambda\gamma^{p}\theta)(\theta\gamma_{mnp}\theta)\end{array}\right\} .\label{eq:PScohomology}
\end{equation}
At ghost number zero it contains only the identity operator. At ghost
number one, $A_{\alpha}$ denotes the super Maxwell superfield and
BRST closedness implies the equation of motion
\begin{equation}
D_{\alpha}\gamma_{mnpqr}^{\alpha\beta}A_{\beta}=0,
\end{equation}
where $D_{\alpha}$ is the superderivative given by
\begin{equation}
D_{\alpha}\equiv\partial_{\alpha}+(\gamma^{m}\theta)_{\alpha}\partial_{m}.\label{eq:superderivative}
\end{equation}
BRST-exact states account for the gauge transformation $\delta A_{\alpha}=D_{\alpha}\Lambda$,
where $\Lambda$ is a superfield parameter. At ghost number two, $A_{\alpha\beta}$
denotes the superfield containing the super Maxwell antifields, and
at ghost number three the only cohomology element is the so-called
pure spinor measure factor.

Finally, the cohomology of the BRST charge \eqref{eq:QPS+U1} is given
by a direct product of the cohomology of its two pieces, consisting
of a doubling of the pure spinor superparticle cohomology: one sector
completely independent of the $U(1)$ variables $\{\gamma,\beta,\Omega,\bar{\Omega}\}$
and one sector linear in $\Omega$, both with the same physical content
of \eqref{eq:PScohomology}.

\section{Summary and final remarks\label{sec:final}}

The BRST quantization of the massless particle action subjected to
the constraint
\begin{equation}
\bar{p}^{a}+\gamma^{ab}p_{b}=0,\label{eq:MasslessSolutionFR}
\end{equation}
was shown here to give rise to an interesting structure (subsection
\ref{subsec:cohomology}). In particular, its ghost number zero cohomology
is described by the wave function
\begin{equation}
F=F(\mathbb{X}^{a},\gamma^{ab}).
\end{equation}
which was used in \cite{Berkovits:2004bw} as a generating function
for massless solutions of field equations of spin $\tfrac{N}{2}$,
\emph{cf}. equation \eqref{eq:masslesssolutions}.

Equation \eqref{eq:MasslessSolutionFR} was shown in subsection \ref{subsec:projective-twistor}
to be equivalent to the twistor-like constraint
\begin{equation}
(\gamma^{m}\lambda)_{\alpha}P_{m}=0,\label{eq:BerkovitsTwistorLikeFR}
\end{equation}
suggested as the fundamental gauge structure behind the pure spinor
superparticle \cite{Berkovits:2011gh,Berkovits:2015yra}. However,
\emph{cf}. subsection \ref{subsec:covariant}, it does not seem to
be possible to obtain the superparticle solely from gauge fixing the
action \eqref{eq:Berkovitsaction} proposed in \cite{Berkovits:2015yra}.

With the introduction of a partial worldline supersymmetry connecting
the spacetime spinors (section \ref{sec:PSsuperparticle}), the BV
quantization of the new model leads to the gauge fixed action
\begin{equation}
S_{\tiny{PS}}=\int d\tau\{P_{m}\dot{X}^{m}-\tfrac{1}{2}P_{m}P^{m}+p_{\alpha}\dot{\theta}^{\alpha}+w_{\alpha}\dot{\lambda}^{\alpha}+\bar{\Omega}\dot{\Omega}+\beta\dot{\gamma}\},\label{eq:FINALPSFR}
\end{equation}
expressed in terms of the unconstrained spinors $\theta^{\alpha}$
and $p_{\alpha}$. Its BRST charge is given by the usual pure spinor
BRST charge plus a $U(1)$ decoupled sector, \emph{cf}. equation \eqref{eq:QPS+U1}.

It is interesting to observe the connections between different (super)particle
descriptions. With the addition of a \emph{partial} worldline supersymmetry
to Berkovits' twistor-like constraint, spacetime supersymmetry emerges,
giving rise to the pure spinor superparticle in $D=10$, which has
the same physical spectrum of the Brink-Schwarz superparticle \cite{Brink:1981nb}.
In turn, as demonstrated in \cite{Sorokin:1988nj}, the Brink-Schwarz
superparticle and the spinning particle \cite{Brink:1976sz}, with
worldline supersymmetry, are classically equivalent.

There are several directions to be explored that could help to understand
these connections in more detail. For example, the higher ghost number
cohomology of \eqref{eq:BRSTcharge} was not determined here but it
could be interesting to explore its field content using the techniques
of \cite{Berkovits:2004bw} with a generating function with an intrinsic
gauge structure. Concerning the action \eqref{eq:PSunfixed}, its
covariant quantization in even dimensions might lead to (alternative)
superparticle descriptions in $D<10$. Another interesting idea is
to enhance it with full worldline supersymmetry and to analyze the
possible connections between the pure spinor superparticle and an
extended version of the spinning particle. This idea is supported
by a recent work relating the Ramond-Neveu-Schwarz and pure spinor
superstrings \cite{Berkovits:2016xnb}.

Probably the most interesting direction to be investigated is the
extension of the results of section \ref{sec:PSsuperparticle} to
the worldsheet, trying to understand the gauge fixing leading to the
pure spinor superstring \cite{Berkovits:2000fe}. The results of \cite{Berkovits:2015yra}
are likely to be completed using the model presented here. Partial
computations already reveal the same pattern \cite{Jusinskas:2019a}, and the string version
of the constraints \eqref{eq:PSconstraints} seem to lead to a consistent
generalization of the gauge fixed action \eqref{eq:PSfinal-scale}.
As already pointed out in \cite{Berkovits:2015yra,Berkovits:2014aia},
reparametrization symmetry is redundant with the implementation of
the twistor-like constraint, but at the worldsheet level this is more
subtle. It would be interesting to have a (worldsheet) covariant derivation
of all these results. Finally, the idea introduced in \cite{Berkovits:2015yra}
that the Green-Schwarz and the pure spinor formalisms are but different
gauge fixings of the same master action is worth investigating, although
this does not look so straightforward considering the constraints
\eqref{eq:PSconstraints}.

\

\textbf{Acknowledgments:} I would like to thank Thales Azevedo, Nathan
Berkovits, Kara Farnsworth and Ond\v{r}ej Hul\'ik for their comments
and discussions, and especially Andrei Mikhailov for the valuable
suggestions. This research has been supported by the Grant Agency
of the Czech Republic, under the grant P201/12/G028.

\end{document}